\documentclass[conference]{IEEEtran}
\IEEEoverridecommandlockouts
\usepackage[utf8]{inputenc} 
\usepackage{amsmath,amssymb,amsfonts} 
\usepackage{graphicx} 
\usepackage{cite} 
\usepackage{url} 
\usepackage{booktabs} 
\usepackage{algorithmicx} 
\usepackage{algpseudocode} 
\usepackage{subcaption} 
\usepackage{multirow} 
\usepackage{array} 
\usepackage{textcomp}
\usepackage{xcolor}
\def\BibTeX{{\rm B\kern-.05em{\sc i\kern-.025em b}\kern-.08em
    T\kern-.1667em\lower.7ex\hbox{E}\kern-.125emX}}
\begin{document}

\title{The Trip to ZigBee Backscatter across a Decade—a Systematic Review}

\author{\IEEEauthorblockN{Yang Liu}
\IEEEauthorblockA{\textit{School of Computer Science} \\
\textit{University of Science and Technology of China}\\
Hefei, China \\
current@mail.ustc.edu.cn}
}

\maketitle
\begin{abstract}
The field of backscatter communication has undergone a profound transformation, evolving from a niche technology for radio-frequency identification (RFID) into a sophisticated paradigm poised to enable a truly battery-free Internet of Things (IoT). This evolution is built upon a deepening understanding of the fundamental principles governing these ultra-low-power links. Modern backscatter systems are no longer simple reflectors of continuous waves but are increasingly designed to interact with complex, data-carrying ambient signals from ubiquitous sources like WiFi, ZigBee, and cellular networks. This review systematically charts the journey of ambient backscatter, particularly focusing on its interaction with ZigBee and other commodity wireless protocols over the last decade. We analyze the progression from foundational proof-of-concept systems that established productive backscatter to modern high-throughput, concurrent, and cross-technology communication architectures. Key advancements in fine-grained modulation, robust synchronization, cross-technology physical layer emulation, and multi-tag coordination are detailed. A comparative analysis of state-of-the-art systems highlights the core trade-offs between performance metrics like data rate and range, power consumption, and compatibility with commodity hardware. Finally, we synthesize the primary challenges—including networking scalability, security vulnerabilities, the near-far problem, and practical deployment hurdles—and outline future research directions, such as integration with Reconfigurable Intelligent Surfaces (RIS) and 6G networks, that promise to further expand the capabilities of this transformative technology.
\end{abstract}


\section{Introduction}

\subsection{Background}
The proliferation of the Internet of Things (IoT) has created an unprecedented demand for low-power, low-cost, and ubiquitous wireless communication. At the heart of this demand lies the challenge of powering billions, and potentially trillions, of connected devices. Traditional battery-powered solutions are often impractical due to the immense cost and logistical burden of battery replacement and disposal, not to mention the environmental impact \cite{jiang2023backscatter}. This has fueled intensive research into alternative communication paradigms that can operate with minimal or zero battery power, a concept often termed the Battery-Free Internet of Things (BF-IoT).

Two key technologies have emerged as central to this pursuit: the ZigBee protocol and backscatter communication. ZigBee, based on the IEEE 802.15.4 standard, is a widely adopted protocol for low-power, low-data-rate wireless personal area networks (WPANs). Its characteristics—including mesh networking, simple protocol stack, and notably low active and sleep power consumption—make it ideal for applications like smart homes, industrial control, and environmental monitoring \cite{dementyev2013power}. However, even with its optimizations, active ZigBee radios still consume milliwatts of power during transmission, which can be a significant drain for devices intended to last for years without intervention.

Backscatter communication offers a more radical approach to power reduction. Instead of actively generating their own radio waves, backscatter devices (or "tags") communicate by reflecting and modulating existing radio frequency (RF) signals present in the environment \cite{liu2013ambient}. This passive approach, where the power-hungry components of a traditional radio (like power amplifiers and oscillators) are eliminated, allows for devices that consume orders of magnitude less power—microwatts instead of milliwatts. This fundamental efficiency enables the vision of truly battery-free IoT devices that can be powered solely by harvesting energy from the very RF signals they use for communication \cite{huynh2018ambient}.

\subsection{Motivation and Significance}
The motivation for combining ZigBee and backscatter communication is compelling and synergistic. By enabling passive tags to communicate using ambient ZigBee signals (or, more powerfully, to generate ZigBee-compatible signals by backscattering other ambient sources like WiFi), we can massively expand the IoT ecosystem. This fusion allows for the deployment of vast networks of simple, disposable, and battery-free sensors that can seamlessly integrate with existing ZigBee infrastructure. The application potential is immense, spanning from smart packaging that reports its status, infrastructure health monitoring sensors embedded in concrete, and dense sensor deployments in smart agriculture, to implantable medical devices that can communicate directly with commodity gateways like smartphones and smartwatches \cite{jiang2023backscatter, iyer2016inter}. This approach promises to unlock the full potential of a pervasive and sustainable IoT, breaking the energy barrier that has long constrained its growth.

\subsection{Review Objectives and Scope}
This paper aims to provide a systematic review of the evolution of ZigBee and related ambient backscatter communication technologies over the past decade (approximately 2015-2025). Our objective is to chart the journey from early foundational concepts that proved the feasibility of "productive" backscatter to the latest high-performance systems capable of multi-megabit throughput and cross-protocol translation. We identify the key technological advancements, synthesize and compare the performance metrics reported across the literature, and discuss the persistent challenges and promising future research directions. The scope is primarily focused on systems that leverage or create commodity-compatible signals—particularly ZigBee, WiFi, and Bluetooth—as these represent the most practical path toward widespread adoption and a truly interoperable IoT.

\subsection{Paper Structure}
This review is organized as follows. Section \ref{sec:foundations} details the foundational principles of modern backscatter systems, including channel characteristics and the key innovation of codeword translation. Section \ref{sec:throughput} charts the quest for higher throughput, analyzing the move to finer-grained modulation schemes. Section \ref{sec:heterogeneous} explores developments enabling communication in the heterogeneous IoT, focusing on cross-technology communication and concurrency. Section \ref{sec:techniques} delves into the key techniques and implementations covering system architectures, tag design, and receiver signal processing. Section \ref{sec:synthesis} provides a comprehensive synthesis, including a comparative performance evaluation, a discussion of core challenges and limitations. Finally, Section \ref{sec:conclusion} concludes the paper with a summary of the field's progress and an outlook on future directions.

\section{Foundational Principles of Modern Backscatter}
\label{sec:foundations}

The evolution of backscatter communication from a simple RFID-like technology to a sophisticated paradigm capable of interacting with complex ambient signals is built upon a deeper understanding of its underlying principles. This section establishes the theoretical and practical bedrock of this new era of ambient backscatter, analyzing the physical channel, the critical interplay between performance and power, and the pivotal innovation that enabled backscatter to leverage productive, data-carrying wireless traffic.

\subsection{The Backscatter Channel: Theory and Practical Limits}
An ambient backscatter system involves three entities: a legacy transmitter (LT) providing the ambient RF signal (e.g., a WiFi AP), a backscatter device (BD) or tag, and a backscatter receiver (BR) \cite{ye2020outage}. The communication is governed by three distinct wireless links: the forward link (LT-to-BD), the backscatter link (BD-to-BR), and a direct interference link (LT-to-BR) \cite{qian2017noncoherent}. Performance is often analyzed assuming a frequency-flat, block-fading channel model \cite{ye2020outage}.

A critical metric is the outage probability. An outage occurs if the energy harvested by the tag is insufficient ($E_k < P_{c,k}T$) or the Signal-to-Interference-plus-Noise Ratio (SINR) at the receiver is too low ($\gamma_k^b < \gamma_{th}^b$) \cite{ye2020outage}. A key finding is that the co-channel interference from the direct LT-to-BR link is a dominant limiting factor, leading to "outage saturation." Beyond a certain point, increasing the LT's power provides diminishing returns, as the interference floor rises with the signal strength \cite{ye2020outage}. This reveals a fundamental performance ceiling.

Furthermore, obtaining complete Channel State Information (CSI) is often infeasible, necessitating noncoherent detection schemes. Seminal work derived the performance of both an optimal Maximum Likelihood (ML) detector and a lower-complexity joint-energy detector \cite{qian2017noncoherent}. These noncoherent detectors often exhibit an "error floor" at high SNR, where the Bit Error Rate (BER) flattens out, not due to thermal noise, but due to channel characteristics \cite{qian2017noncoherent}. This again highlights that performance gains must come from more sophisticated system design. The non-linearity of real-world energy harvesters (EH) further complicates this; assuming a simple linear EH model leads to an overly optimistic assessment of system performance \cite{ye2020outage}.

\subsection{Codeword Translation: The Gateway to Productive Backscatter}
A pivotal innovation that unlocked the potential of ambient backscatter is "codeword translation" \cite{zhang2017freerider}. This technique allows a tag to operate using a *productive* ambient signal—a standard, data-carrying transmission from a commodity device like a WiFi router—rather than a simple, non-productive continuous wave.

The core concept is to transform a valid codeword from the ambient signal's codebook into another valid codeword from the *same* codebook during reflection. A "codeword" here is a physical layer signal symbol (e.g., a phase-modulated state). A "codebook" is the complete set of valid symbols defined by a protocol. The tag accomplishes this by manipulating the signal's amplitude, phase, or frequency. In a simple binary scheme, to send a '1', a tag might introduce a $180^\circ$ phase shift to the incoming codeword; for a '0', it would apply no shift \cite{zhang2017freerider}.

This technique necessitates a bistatic, dual-receiver setup. One receiver decodes the original, unmodified signal. A second receiver, often on an adjacent channel, decodes the backscattered signal. The tag's data is recovered by comparing the two decoded bit streams, typically via a bitwise XOR operation \cite{zhang2017freerider, zhao2018xtandem}. This architecture, foundational to pioneering systems like FreeRider \cite{zhang2017freerider}, proved that backscatter could coexist with and leverage active wireless traffic, paving the way for the integrated single-receiver systems that followed.

\subsection{The Symbiotic Challenge of Synchronization and Power}
High-performance backscatter demands precise synchronization with the incoming carrier, which requires power-intensive, high-bandwidth processing that runs counter to the ultra-low-power ethos of the technology \cite{dunna2021syncscatter}. Early systems used simple energy detection, achieving only coarse synchronization and thus limiting data rates \cite{liu2013ambient}.

The SyncScatter system provided a canonical solution through a two-stage hierarchical wake-up and synchronization protocol, reframing synchronization as a power-managed, event-driven process \cite{dunna2021syncscatter}.
\begin{itemize}
    \item \textbf{Stage 1: Low-Power Wake-up.} A passive, low-bandwidth energy detector, consuming single-digit microwatts, monitors for a pre-specified wake-up signature (e.g., a sequence of WiFi packets with specific lengths).
    \item \textbf{Stage 2: High-Precision Synchronization.} Once awakened, the tag briefly activates a higher-power, higher-bandwidth active RF amplifier for a few microseconds—just long enough to achieve precise symbol-level synchronization (e.g., within 150 ns for WiFi).
\end{itemize}
This aggressive duty-cycling of power-intensive components is the key to minimizing average power consumption while achieving high performance when needed. This design was born from a holistic link budget analysis that considered FCC power limits, receiver sensitivity, path loss, and BER degradation from synchronization errors, establishing a target sensitivity and demonstrating that improving it further yields diminishing returns \cite{dunna2021syncscatter}. This cross-layer optimization approach is a crucial principle for nearly all high-performance, low-power wireless systems.

\section{The Quest for Higher Throughput}
\label{sec:throughput}

The primary driver of innovation in backscatter has been the quest for higher data rates. Early systems, with throughputs in the kilobits-per-second range, were insufficient for a data-rich IoT. The journey to megabit-per-second backscatter is marked by a clear trend: moving from coarse-grained modulation of entire symbols towards fine-grained manipulation of the signal's fundamental components.

\subsection{From Symbols to Samples: A Leap in Granularity for OFDM}
Early attempts to backscatter Orthogonal Frequency Division Multiplexing (OFDM) signals, the foundation of WiFi, faced a formidable obstacle. When a tag introduced a phase shift at the symbol level to encode data, the WiFi receiver's phase error correction algorithm—which uses pilot subcarriers to cancel channel-induced phase errors—would interpret the tag's modulation as just another error and "correct" it, erasing the data \cite{liu2021verification}.

The breakthrough came with TScatter, which introduced *sample-level* modulation \cite{liu2021verification}. Instead of one phase shift per $4~\mu s$ symbol, the TScatter tag toggles its RF switch at the 20 MHz WiFi sample rate (every 50 ns), introducing a unique phase shift on *each individual sample*. The key insight was that this would cause the phase offsets on data-carrying subcarriers to differ from those on pilot subcarriers. This breaks the receiver's assumption of a common phase error. As a result, the receiver corrects the actual channel error using the pilots but leaves the tag's data-bearing phase information on the data subcarriers intact.

This innovation, combined with a demodulation model that estimates tag data by minimizing the Euclidean distance between received subcarrier values and their expected QAM constellation points, enabled a massive leap in performance. TScatter demonstrated throughputs up to 13.63 Mbps, three orders of magnitude higher than previous systems, while remaining compatible with unmodified commodity WiFi receivers \cite{liu2021verification}.

\subsection{Chip-Level Modulation: Unleashing ZigBee's Potential}
A parallel evolution occurred in ZigBee backscatter. ZigBee uses Direct-Sequence Spread Spectrum (DSSS), where every 4 data bits are mapped to a 32-chip pseudo-random sequence \cite{wang2023embracing}. Early systems like FreeRider used coarse-grained, symbol-level modulation for robustness, applying the same phase shift across eight consecutive symbols to encode a single tag bit. This came at the cost of a throughput reduction of over 32 times compared to active ZigBee \cite{zhang2017freerider}.

The breakthrough, pioneered by systems like ChipScatter and EchScatter, was to modulate at the level of individual chips within a single symbol \cite{wang2023poster, li2023echscatter}. The tag applies a carefully designed 32-chip phase modulation sequence to the 32 chips of the incoming symbol. This "enriches" codeword translation: the goal is to transform any of the 16 possible incoming ZigBee symbols into any of the other 15.

To achieve this, EchScatter designed 16 unique 32-chip phase modulation sequences, each corresponding to a 4-bit tag data value ($2^4=16$). When a tag sends a 4-bit value, it applies the corresponding phase sequence to the incoming ZigBee symbol. The resulting backscattered chip sequence, when processed by the commodity ZigBee receiver's minimum Hamming distance decoder, is uniquely decoded as one of the 16 possible output symbols. By comparing the original and backscattered symbols, a look-up table recovers the 4-bit tag data \cite{li2023echscatter}.

This fine-grained control allows a single ZigBee symbol to carry 4 bits of tag data, a dramatic improvement over FreeRider's 1 bit per 8 symbols. This catapulted ZigBee backscatter throughput to rates comparable to active ZigBee (e.g., 247 kbps), a 32-fold increase over symbol-level methods \cite{li2023echscatter}.

\subsection{Intelligent Rate Adaptation for Mobile Backscatter Networks}
Achieving high throughput also requires intelligently adapting the transmission rate to the dynamic wireless channel, especially in mobile scenarios. Early rate adaptation schemes for backscatter (e.g., Blink, CARA) relied on a static, pre-trained 2D map, selecting the optimal rate based on Received Signal Strength Indicator (RSSI) and packet loss rate \cite{gong2018mobirate, wang2016towards}. This approach is brittle, as the maps are highly hardware-dependent; a map trained for one tag model performs poorly on another \cite{gong2018mobirate}.

MobiRate overcomes this by replacing the static map with a dynamic framework that leverages PHY-layer mobility information \cite{gong2018mobirate}. It introduces three key components:
\begin{enumerate}
    \item \textbf{Velocity-Based Loss Rate Estimation:} Packet loss is not static. MobiRate provides a more accurate estimate of channel quality by re-weighing loss statistics based on the tag's velocity, derived from the Doppler shift in the signal's phase.
    \item \textbf{Mobility-Assisted Probing Trigger:} Channel probing (testing different rates) is a source of overhead. MobiRate reduces this by using the tag's location and direction of movement to eliminate unnecessary probes.
    \item \textbf{Selective, Collision-Free Probing:} To address probe collisions in multi-tag environments, MobiRate repurposes the `SELECT` command of the ISO 18000-6C standard. This command, normally used for inventorying a subset of tags, is used to enable collision-free, point-to-point probing of individual tags.
\end{enumerate}
By combining these techniques, MobiRate achieves throughput gains of up to 3.8x over previous systems, demonstrating the superiority of adaptive, PHY-aware systems in real-world mobile environments \cite{gong2018mobirate}.

\section{The Heterogeneous IoT: Cross-Technology and Coexistence}
\label{sec:heterogeneous}
The modern IoT is a dense, heterogeneous ecosystem where devices using different wireless standards like WiFi, ZigBee, and Bluetooth must coexist and, ideally, cooperate. Backscatter communication has emerged as a powerful technology to bridge these disparate worlds.

\subsection{Cross-Technology PHY Emulation}
The ultimate goal of Cross-Technology Communication (CTC) is to enable a device using one protocol to generate a signal that can be directly decoded by an unmodified commodity receiver of a different protocol. This requires physical-layer (PHY) emulation, where the tag acts as an on-the-fly protocol translator.

Early work like Interscatter demonstrated transforming BLE transmissions into standards-compliant WiFi and ZigBee signals \cite{iyer2016inter}. This involved forcing the BLE transmitter to emit a simple, non-productive single-tone carrier, then using a sophisticated backscatter modulator to perform single-sideband modulation, impressing the target protocol's waveform onto the carrier.

Subsequent systems refined this. BumbleBee showed it was possible to use a *productive* BLE carrier, arguing that the tag's modulation could be made dominant enough to overwrite the BLE signal's information from the perspective of a less sensitive, narrowband ZigBee receiver \cite{xu2023bumblebee}.

Perhaps the most elegant solution was BlueBee, which achieved PHY emulation purely through payload manipulation \cite{jiang2017bluebee}. Its key insight was that by carefully crafting the bit patterns in a standard BLE packet's payload, the resulting GFSK-modulated RF waveform could be made to physically resemble the waveform of a legitimate OQPSK-modulated ZigBee packet. The inherent error tolerance of the ZigBee receiver's DSSS design was sufficient to correctly decode this imperfect but recognizable emulated packet. This was achieved without modifying the BLE transmitter's firmware, demonstrating true transparency \cite{jiang2017bluebee}.

\subsection{Concurrent Transmissions: Resolving Collisions in Dense Networks}
As IoT deployments become denser, packet collisions become a primary bottleneck \cite{kong2015mzig, wang2019ppm}. A significant body of research has focused on physical-layer techniques for multi-packet reception (MPR), allowing a receiver to decode multiple, overlapping packets. Several approaches for ZigBee have emerged:

\textbf{Interference Cancellation via Time Offsets (mZig/CmZig):} These systems exploit the fact that collided packets almost always arrive with a slight time offset. By oversampling the signal, the receiver can identify a few collision-free samples at the start of the first packet's first chip. Leveraging the known half-sine pulse shape of a ZigBee chip, the receiver reconstructs the entire chip's waveform, subtracts it from the composite signal, and repeats the process for the next packet \cite{kong2015mzig, wang2023embracing}. CmZig refines this by incorporating channel estimation to more accurately model each transmitter's signal \cite{wang2023embracing}.

\textbf{Reference-Based Decoding (PPM):} The Preamble and Postamble-based MPR (PPM) system requires the transmitter to attach a short, known postamble to each packet. The receiver then has known, collision-free reference chips at both the beginning (preamble) and end (postamble) of the collided segment. These are used to construct a library of ideal waveforms for all possible overlap combinations, against which the collided signal is compared \cite{wang2019ppm}.

\textbf{Orthogonal Waveforms (OrthZig):} To avoid complex cancellation or waveform construction, OrthZig assigns mutually orthogonal spreading codes (Walsh codes) to different transmitters. The receiver can separate the linearly combined signals by correlating the composite signal with each transmitter's known orthogonal waveform. This offers high-precision resolution with low computational complexity \cite{wang2023orthzig}.

\subsection{Extending the Network Fabric: Multi-Hop Backscatter}
Single-hop backscatter range is limited by the "doubly near-far" problem, where the signal suffers path loss on both the forward (LT-to-BD) and backscatter (BD-to-BR) links \cite{huynh2018ambient, jiang2023backscatter}. Multi-hop communication, where tags relay signals, is a classic solution to extend network coverage.

X-Tandem was the first to demonstrate a practical multi-hop backscatter architecture compatible with commodity WiFi \cite{zhao2018xtandem}. It is built on two key innovations:
\begin{enumerate}
    \item \textbf{Analog Forwarding:} Instead of decode-and-forward, tags in X-Tandem perform analog forwarding. A relay tag receives the analog waveform from a previous-hop tag and, without decoding, simultaneously re-modulates it with its *own* data before reflecting it onward.
    \item \textbf{Multiple Frequency Shifts (MFS):} To prevent interference between hops, X-Tandem uses MFS. The first tag receives the original signal at $f_c$ and backscatters it at $f_c+f_1$. The second tag receives this and re-transmits at $f_c+f_1+f_2$. The final receiver listens only at the final frequency, isolating the complete multi-hop packet.
\end{enumerate}
This allows a single WiFi packet to travel through a chain of tags, accumulating data from each before being decoded by a single commodity WiFi receiver \cite{zhao2018xtandem}.

\section{Key Techniques and Implementations}
\label{sec:techniques}

\subsection{System Architectures}
\subsubsection{Monostatic vs. Bistatic}
Backscatter systems are broadly categorized into two architectures. In a \textbf{monostatic} system, a single device acts as both the RF source (or "exciter") and the receiver. This is the classic architecture used in commercial RFID readers. It is simpler to deploy but suffers from strong self-interference, as the reader's own powerful transmission can easily overwhelm the tag's faint reflection.

In a \textbf{bistatic} architecture, the RF source and the receiver are separate entities. This is the dominant architecture in ambient backscatter research. For example, in FreeRider \cite{zhang2017freerider}, the exciter is a commodity WiFi AP, and the receiver is a second commodity WiFi device. This separation provides spatial diversity and helps mitigate the direct-path interference, but it introduces the complexity of coordinating three separate devices.

\subsubsection{Ambient Backscatter}
This is a specific form of bistatic backscatter where the RF source is an "ambient" transmitter that is not part of the backscatter system itself, such as a public TV tower \cite{liu2013ambient}, a cellular base station \cite{chi2020leveraging}, or a nearby WiFi access point \cite{zhang2017freerider}. The key advantage is that it eliminates the need to deploy a dedicated power-hungry exciter, allowing tags to communicate opportunistically using signals that are already pervasive in the environment. This is the most promising approach for enabling a truly ubiquitous and battery-free IoT.

\subsection{Tag Design}
\subsubsection{Antenna and RF Front-end}
The tag's front-end is remarkably simple. It consists of an antenna connected to an RF switch (typically a single transistor). The core principle of backscatter is impedance mismatch. When the switch is in one state (e.g., "off"), the antenna's impedance is matched to the load, and it absorbs maximum power from the incident RF wave. When the switch is in the other state ("on"), the impedance is mismatched, causing the antenna to reflect the incident wave. By toggling this switch, the tag modulates the reflected signal \cite{huynh2018ambient}. The evolution of tag design has focused on optimizing antenna efficiency and minimizing the power required to operate this switch.

\subsubsection{Modulation Schemes}
The data from the tag is encoded by the pattern of toggling the RF switch.

\textbf{On-Off Keying (OOK):} This is the simplest scheme, where the tag reflects to send a '1' and absorbs (does not reflect) to send a '0'. This is used in many early and simple backscatter systems.

\textbf{Phase-Shift Keying (PSK):} By using a more complex switching network, the tag can introduce a phase shift onto the reflected signal. Binary PSK (BPSK), where a $180^\circ$ phase shift is used to encode data, is common. This is the basis for codeword translation in systems like FreeRider \cite{zhang2017freerider}.

\textbf{Frequency-Shift Keying (FSK):} By toggling the switch at a specific frequency $f_m$, the tag can shift the frequency of the carrier signal, creating sidebands at $f_c \pm f_m$. This is used in systems like X-Tandem to separate signals from different hops \cite{zhao2018xtandem}. More recent work has proposed frequency-phase shift (FPS) modulation, a fine-grained technique that creates a continuous phase shift to suppress spectrum sidelobes, improving spectral efficiency \cite{xu2022enabling}.

More advanced systems like EchScatter \cite{li2023echscatter} use complex, pre-designed sequences of phase shifts applied at the chip level to achieve higher-order modulation.

\subsubsection{Energy Harvesting}
A key capability of passive tags is harvesting energy from the incident RF signal to power their own circuitry. The same antenna used for communication receives RF energy, which is fed to a rectifier circuit (typically using Schottky diodes) to convert it into a usable DC voltage. The efficiency of this process is a critical bottleneck, as typical RF energy densities are very low. As noted earlier, real-world energy harvesters exhibit significant non-linearity, a factor that must be considered for accurate performance analysis \cite{ye2020outage}.

\subsection{Receiver Design and Signal Processing}

\subsubsection{Interference Cancellation}
The single greatest challenge at the receiver is canceling the overwhelmingly strong direct-path interference from the RF source. The receiver sees a composite signal containing the powerful signal directly from the source and the extremely weak, data-carrying reflection from the tag. Early systems like FreeRider \cite{zhang2017freerider} used a second receiver to obtain a clean copy of the source signal, which could then be subtracted from the composite signal. More advanced systems aim for single-receiver designs, employing sophisticated analog or digital cancellation techniques to isolate the backscattered signal.

\subsubsection{Decoding Algorithms}
Once the interference is canceled, the backscattered signal must be decoded. For simple OOK, this can be done with a simple energy detector. For PSK-based systems using codeword translation, the process is more complex. The receiver decodes the full packet and then compares the resulting bitstream to the original bitstream (obtained via a second receiver or known a priori) to extract the tag's data, often via a simple XOR operation \cite{zhang2017freerider}. In chip-level modulation systems like EchScatter, the receiver uses the known properties of the ZigBee DSSS decoder to determine which 4-bit data value the tag sent based on which symbol transformation occurred \cite{li2023echscatter}.

\section{Synthesis, Challenges, and Future Outlook}
\label{sec:synthesis}

\begin{table*}[t]
\centering
\caption{Performance and Architectural Comparison of Modern Backscatter Systems (Reformatted for Readability)}
\label{tab:comparison}
\footnotesize 
\renewcommand{\arraystretch}{1.2} 
\begin{tabular}{@{}l l l l@{}}
\toprule
\textbf{System} & \textbf{Excitation Signal / Data Rate} & \textbf{Target Protocol / Range} & \textbf{Key Innovation / Tag Power (µW) and Limitations} \\
\midrule

\textbf{FreeRider \cite{zhang2017freerider}} & \textbf{Excitation:} WiFi, ZigBee, BLE & \textbf{Target:} Same as Excitation & \textbf{Innovation:} Symbol-level codeword translation \\
& \textit{Rate:} $\sim$60 Kbps (WiFi) & \textit{Range:} up to 42m (WiFi) & \textit{Power:} $\sim$30 µW. \textit{Limits:} Requires dual-receiver setup \\
\addlinespace[4pt]

\textbf{Interscatter \cite{iyer2016inter}} & \textbf{Excitation:} BLE (single-tone) & \textbf{Target:} WiFi, ZigBee & \textbf{Innovation:} Single-sideband backscatter \\
& \textit{Rate:} 2--11 Mbps (WiFi) & \textit{Range:} N/S & \textit{Power:} 28 µW. \textit{Limits:} Needs non-productive carrier \\
\addlinespace[4pt]

\textbf{BlueBee \cite{jiang2017bluebee}} & \textbf{Excitation:} BLE (productive) & \textbf{Target:} ZigBee & \textbf{Innovation:} Cross-tech PHY emulation \\
& \textit{Rate:} 225 Kbps & \textit{Range:} $\sim$10m & \textit{Power:} N/S. \textit{Limits:} Bandwidth mismatch limits performance \\
\addlinespace[4pt]

\textbf{X-Tandem \cite{zhao2018xtandem}} & \textbf{Excitation:} WiFi (productive) & \textbf{Target:} WiFi & \textbf{Innovation:} Multi-hop analog forwarding, MFS \\
& \textit{Rate:} up to 200 bps & \textit{Range:} up to 8m (2-hop) & \textit{Power:} 14,200 µW (FPGA). \textit{Limits:} Very low throughput, FPGA power \\
\addlinespace[4pt]

\textbf{SyncScatter \cite{dunna2021syncscatter}} & \textbf{Excitation:} WiFi (productive) & \textbf{Target:} WiFi & \textbf{Innovation:} Hierarchical wake-up \& sync \\
& \textit{Rate:} 500 Kbps & \textit{Range:} 30+ m & \textit{Power:} 30 µW. \textit{Limits:} Custom ASIC needed for low power \\
\addlinespace[4pt]

\textbf{TScatter \cite{liu2021verification}} & \textbf{Excitation:} WiFi (OFDM) & \textbf{Target:} WiFi & \textbf{Innovation:} Sample-level modulation \\
& \textit{Rate:} \textbf{13.63 Mbps} & \textit{Range:} up to 160 ft & \textit{Power:} 30.2 µW. \textit{Limits:} Implemented on SDR, not commodity \\
\addlinespace[4pt]

\textbf{EchScatter \cite{li2023echscatter}} & \textbf{Excitation:} ZigBee (productive) & \textbf{Target:} ZigBee & \textbf{Innovation:} Chip-level modulation \\
& \textit{Rate:} \textbf{247 Kbps} & \textit{Range:} up to 20m & \textit{Power:} 280,000 µW (FPGA). \textit{Limits:} High power on prototype (FPGA) \\
\addlinespace[4pt]

\textbf{BumbleBee \cite{xu2023bumblebee}} & \textbf{Excitation:} BLE (productive) & \textbf{Target:} ZigBee & \textbf{Innovation:} Dominant tag data overwrite \\
& \textit{Rate:} 218 Kbps & \textit{Range:} up to 20m & \textit{Power:} N/S. \textit{Limits:} Relies on receiver error tolerance \\
\addlinespace[4pt]

\textbf{Multiscatter \cite{gong2020multiprotocol}} & \textbf{Excitation:} WiFi, BLE, ZigBee & \textbf{Target:} Same as Excitation & \textbf{Innovation:} Multiprotocol ID, Overlay Mod. \\
& \textit{Rate:} 278 Kbps (agg.) & \textit{Range:} 20--28m & \textit{Power:} 2,000 µW (opt.). \textit{Limits:} Requires custom carrier \\

\bottomrule
\end{tabular}
\end{table*}

\subsection{Performance Metrics Comparison}
To distill the dense technical information from the surveyed literature, Table \ref{tab:comparison} presents a comparative analysis of key backscatter systems. This table highlights architectural choices, key innovations, and reported performance metrics, allowing for a direct comparison of their capabilities and limitations. It visualizes the design space, illustrating the trade-offs between throughput, range, power consumption, and compatibility.

\subsection{Core Challenges}
Synthesizing the literature reveals a set of common, fundamental challenges that researchers in this field continue to grapple with.

\textbf{The Near-Far Problem:} In any multi-tag system, the strong signal from a tag located near the receiver can easily drown out the much weaker signal from a tag that is farther away \cite{huynh2018ambient, jiang2023backscatter}. This is a classic problem in wireless communication that is exacerbated in backscatter due to the passive nature of the tags. Mitigating this is essential for enabling reliable concurrent transmissions in dense networks. Potential solutions include power control where tags adjust their reflection coefficients or advanced receivers using Successive Interference Cancellation (SIC).

\textbf{The Data Rate vs. Range Trade-off:} There is an inherent trade-off between how fast a tag can transmit and how far away it can be. High data rates require more complex modulation and more precise synchronization, which in turn require more power and a stronger incident signal, limiting range \cite{guo2022saiyan}. Low-rate systems like those based on LoRa backscatter can operate at longer distances but are unsuitable for many emerging applications \cite{guo2022saiyan}.

\textbf{Multi-tag Collision:} When multiple tags attempt to backscatter a signal simultaneously, their reflections interfere with each other at the receiver, causing a "collision" where no data can be decoded. While MAC protocols in active networks (like CSMA/CA) address this, designing efficient, ultra-low-power MAC protocols for passive tags is a significant challenge \cite{kong2015mzig}.

\subsection{Research Limitations}
Despite impressive results, much of the existing research has common limitations. A primary one is the reliance on controlled laboratory environments. Performance metrics reported in papers are often achieved under ideal line-of-sight conditions with minimal external interference. The robustness and reliability of these systems in complex, real-world deployments (e.g., a crowded public space or a dynamic industrial setting) remain largely unvalidated. Furthermore, many of the most advanced systems are prototyped on power-hungry FPGAs or SDRs \cite{zhao2018xtandem, li2023echscatter}. The transition to low-cost, ultra-low-power ASICs is a critical but difficult step that is necessary for practical deployment.

\section{Conclusion and Future Directions}
\label{sec:conclusion}
\subsection{Concluding Summary}
The past decade has witnessed a remarkable evolution in ZigBee and ambient backscatter technology. The field has progressed from initial feasibility studies demonstrating basic codeword translation to sophisticated systems capable of high-throughput, cross-technology, and concurrent communication. The core progress has been driven by a move towards finer-grained modulation at the chip and sample level, a deeper, protocol-specific understanding of commodity receiver logic, and the development of novel techniques for synchronization and multi-tag coordination. These advancements have transformed backscatter from a simple RFID-like technology into a versatile communication primitive that is poised to become a cornerstone of the battery-free Internet of Things. The shift from dual-receiver proof-of-concepts to more practical single-receiver designs marks a critical maturation point, signaling a move towards real-world deployability.

\subsection{Future Research Directions}
Based on the challenges and limitations identified, several promising research directions emerge for the future.

\textbf{Security and Privacy:} The passive and open nature of backscatter makes it vulnerable to eavesdropping, jamming, and spoofing attacks. Given the severe power and computational constraints of tags, traditional cryptography is often infeasible. Future work must focus on developing lightweight security mechanisms, potentially leveraging physical-layer properties like channel randomness to create secure, low-power communication links \cite{he2021cross}.

\textbf{Networking and Standardization:} While the physical layer has seen immense innovation, the MAC and networking layers for large-scale backscatter networks are still in their infancy. Designing scalable MAC protocols to manage channel access for thousands of tags is a critical open problem \cite{wang2023orthzig}. Standardization efforts are vital for creating an interoperable ecosystem.

\textbf{Intelligent Surfaces and Beamforming:} New technologies like Reconfigurable Intelligent Surfaces (RIS) offer exciting possibilities. An RIS is a planar surface with many passive elements that can be electronically controlled to reflect RF signals in a specific direction. Integrating RIS with backscatter could allow for intelligent focusing of ambient energy onto tags and steering of backscattered signals towards a receiver, dramatically overcoming path loss and extending range and reliability.

\textbf{Cross-Technology Integration and Coexistence:} Future research will likely deepen the integration with other wireless technologies. This includes not only cross-technology communication but also graceful coexistence with WiFi, 5G, and future 6G networks. This involves designing backscatter systems that can operate robustly in an increasingly crowded spectrum and potentially even leverage the complex signal structures of next-generation cellular networks as a source for high-quality ambient power and carriers.

\bibliographystyle{IEEEtran}
\bibliography{references}

\end{document}